\documentclass[twocolumn,aps,prl,10pt,superscriptaddress]{revtex4-1} %superscriptaddress, groupedaddress,showpacs,floatfix,aps,10pt,amsmath,amssymb,preprintnumbers

\usepackage{graphicx}
\usepackage{bm}
\usepackage{dcolumn}
\usepackage{amssymb}
\usepackage{amsmath}
\usepackage{amsbsy}
\usepackage{amsfonts}
\usepackage{epsfig}
\usepackage{graphicx}
\usepackage{multirow}
\usepackage{stackrel}
\usepackage{paralist}
\usepackage[rgb]{xcolor}
\usepackage[textsize=tiny,textwidth = 1.6 cm]{todonotes}
\usepackage[export]{adjustbox}

\usepackage[percent]{overpic}
\usepackage{hyperref}
\usepackage[caption=false]{subfig}
\usepackage[normalem]{ulem}

\newcommand{\cancel}[1]{}%{\sout{#1}} %{}%
\newcommand{\I}{\mathrm{i}}

\DeclareMathOperator{\sgn}{sgn}

\newcommand{\refEq}[1]{Eq.~(\ref{#1})}
\newcommand{\refFig}[1]{Fig.~\ref{#1}}
\newcommand{\citeRef}[1]{Ref.~[\onlinecite{#1}]}

\definecolor{NewColor}{rgb}{1,0,0}
\definecolor{myRed}{rgb}{1,0,0}
\definecolor{myGreen}{rgb}{0.2,0.6,0.2}
\definecolor{myBlue}{rgb}{0,0,1}

\graphicspath{{./}{./../Graphics/}{./../Graphics/Publication/}}

\newcommand{\CO}[1]{\textcolor{red}{}}

\newlength{\figHeight}
\newlength{\noLengthl}

\usepackage{amsmath,amssymb,amsfonts,graphicx,multirow,color,bm}
\newcommand{\bee}{\begin{eqnarray}}
\newcommand{\ee}{\end{eqnarray}}
\newcommand{\bma}{\begin{pmatrix}}
\newcommand{\ema}{\end{pmatrix}}
\newcommand{\balig}{\begin{align}}
\newcommand{\ealig}{\end{align}}

\newcommand{\ba}{\begin{align}}
\newcommand{\ea}{\end{align}}

\newcommand{\ignore}[1]{}

\def\sgn{\mathop{\textrm{sgn}}}

\newcolumntype{C}[1]{>{\centering\let\newline\\\arraybackslash\hspace{0pt}}m{#1}}

\begin{document}
\setlength{\figHeight}{25ex}

% -------- Title -----------
\title{Bosonic continuum theory of one-dimensional lattice anyons}
\author{Martin Bonkhoff}
\affiliation{Physics Department and Research Center Optimas, Technische Universit{\"a}t Kaiserslautern, 67663 Kaiserslautern, Germany}
\author{Kevin J\"agering}
\affiliation{Physics Department and Research Center Optimas, Technische Universit{\"a}t Kaiserslautern, 67663 Kaiserslautern, Germany}
\author{Sebastian Eggert}
\affiliation{Physics Department and Research Center Optimas, Technische Universit{\"a}t Kaiserslautern, 67663 Kaiserslautern, Germany}
\author{Axel Pelster}
\affiliation{Physics Department and Research Center Optimas, Technische Universit{\"a}t Kaiserslautern, 67663 Kaiserslautern, Germany}
\author{Michael Thorwart}
\affiliation{I. Institut f{\"u}r Theoretische Physik, Universit{\"a}t Hamburg, Jungiusstra{\ss}e 9, 20355 Hamburg, Germany}
\affiliation{The Hamburg Centre for Ultrafast Imaging, Luruper Chaussee 149, 22761 Hamburg, Germany}
\author{Thore Posske}
\affiliation{I. Institut f{\"u}r Theoretische Physik, Universit{\"a}t Hamburg, Jungiusstra{\ss}e 9, 20355 Hamburg, Germany}
\affiliation{The Hamburg Centre for Ultrafast Imaging, Luruper Chaussee 149, 22761 Hamburg, Germany}
\begin{abstract}
Anyons with arbitrary exchange phases exist on 1D lattices in ultracold gases. Yet, known continuum theories in 1D do not match. We derive the continuum limit of 1D lattice anyons via interacting bosons. The theory maintains the exchange phase periodicity fully analogous to 2D anyons. This provides a mapping between experiments, lattice anyons, and continuum theories, including Kundu anyons with a natural regularization as a special case. We numerically estimate the Luttinger parameter as a function of the exchange angle to characterize long-range signatures of the theory and predict different velocities for left- and right-moving collective excitations.
\end{abstract}
\maketitle

Acquiring ``any phase'' when spatially exchanged
\cite{Wilczek1982QuantumMechanicsOfFractionalSpinParticles,LeinaasMyrheim1977TheoryOfIdenticalParticles},
anyons break the dichotomic classification of quantum particles into bosons and fermions. 
Anyons have inspired theoretical and experimental physicists for decades \cite{% 
LeinaasMyrheim1977TheoryOfIdenticalParticles,Wilczek1982QuantumMechanicsOfFractionalSpinParticles,GoldinMenikoffSharp1980ParticleStatisticsFromInducedRepresentationsOfALocalCurrentGroup,BiedenharnLiebSimonWilczek1990TheAncestryOfTheAnyon,Haldane1991FractionalStatisticsInArbitraryDimensionsAGeneralizationOfThePauliPrinciple,khare2005FractionalStatisticsAndQuantumTheory, MooreRead1991NonabelionsInTheFractionalQuantumHallEffect,Laughlin1983AnomalousQuantumHallEffectAnIncompressibleQuantumFluidWithFractionallyChargedExcitations,CaminoZhouWeiGoldman2005RealizationOfALaughlinQuasiparticleInterferometerObservationOfFractionalStatistics,	WilletPfeifferWest2010AlternationAndInterchangeOfEo4andEo2PeriodInterferenceOscillationsConsistentWithFillingFactor5o2NonAbelianQuasiparticles,ZhongXuWangSonGuoLiuXuXuaLuHanPanWang2016EmulatingAnyonicFractionalStatisticalBehaviorInASuperconductingQuantumCircuit}.
Recently, scientific and technological interest increased again because of possible applications of non-abelian anyons in topological quantum computing \cite{Kitaev2006AnyonsInAnExactlySolvableModelAndBeyond,NayakSternFreedmanDasSarma2008NonAbelianAnyonsAndTopologicalQuantumComputation} and the apparent detection of Majorana zero modes \cite{Kitaev2001,	Ivanov2001NonAbelianStatisticsOfHalfQuantumVortices,MourikZuoFrolovPlissardBakkersKouwenhoven2012SignaturesOfMajoranaFermionsInHybridSuperconductorSemiconductorNanowireDevice,KimPalacioMoralesPosskeEtAl2018TowardTailoringMajoranaBoundStatesInArtificiallyConstructedMagneticAtomChainsOnElementalSuperconductors}.
Even though anyons are often considered to be exclusively two-dimensional, 
they have also been discussed
in a wide range of one-dimensional (1D) systems that differ strongly in phenomenology \cite{AliceaFendley2016TopologicalPhasesWithParafermionsTheoryAndBlueprints,DaiXie2009IntermediateStatisticsSpinWaves,Hansson1992DimensionalReductionInAnyonSystems,Ha1995FractionalStatisticsInOneDimensionViewFromAnExactlySolvableModel,Pasquier1994ALectureOnTheCalogeroSutherlandModels,Fisher1997,BatchelorGuanOelkers2006OneDimensionalInteractingAnyonGasLowEnergyPropertiesAndHaldaneExclusionStatistics,BrinkHanssonVasiliev1992ExplicitSolutionToTheNBodyCalogeroProblem,HaoZhangChen2009GroundStatePropertiesOfHardCoreAnyonsIn1DOpticalLattices,ValienteBoseFermidualitiesforarbitraryonedimensionalquantumsystemsintheuniversallowenergyregime}.

In 2D, abelian anyons can exist because the possible continuous 
exchange of two indistinguishable particles provides topological 
nontriviality at a fundamental level.
The wave function of two anyons acquires a phase 
factor $e^{i\theta}$, where the angle $\theta$ specifies the commutation up to a $2\pi$-periodicity. 
In 1D, the situation is more involved because particles cannot directly be exchanged without collision. This is only circumvented by network-like geometries \cite{Alicea2011Majoana1DNetworks}, additional degrees of freedom \cite{Chiu2015MajoranaShuttle}, or measurement-based protocols \cite{Plugge2017MajoranaBoxQubit}. Historically, several notions of one-dimensional anyons have emerged that are a priori disconnected to 2D anyons.
As a fundamental approach, Leinaas and Myrheim have classified indistinguishable 
one-dimensional particles by generalized boundary conditions
in terms of a statistical parameter $-\infty < \eta \leq \infty$ \cite{LeinaasMyrheim1977TheoryOfIdenticalParticles}.
In the mathematically equivalent Lieb-Liniger model of interacting bosons \cite{LiebLiniger1963,PosskeTrauzettelThorwart2016SecondQuantizationOfLeinaasMyrheimAnyonsIn1DAndTheirRElationToTheLLModel}, the statistical parameter is given by the bosonic on-site interaction, and hence {differs strongly} from a $2\pi$-periodic exchange phase $e^{i\theta}$.
Another approach to 
anyonic exchange 
in 1D is to consider a discontinuous jump in the 
wave function when two particles pass each other. This insight was used by Kundu,
who proposed an integrable bosonic model on a continuum 1D system
containing  derivatives and squares of delta-functions \cite{Kundu1999ExactSolutionOfDoubleDeltaFunctionBoseGasThroughAnInteractingAnyonGas} as a model for 1D anyons. 
However,  % $\kappa$ 
%{describing the anyonic exchange} % phase 
 the bosonic form of the Kundu model is not 
$2\pi$-periodic in the statistical parameter either. 

In contrast to the aforementioned continuum theories, a $2\pi$-periodic model of anyons is constructed on a 1D lattice by anyonic creation and annihilation operators with the generalized commutation relation which ensures
\begin{align}
\label{eqnQuantumbrackets}
 {a}_i {a}_j^\dagger - e^{-i\theta \mathrm{sgn}(i-j)} {a}_j^\dagger {a}_i
 = \delta_{i,j},
%  {I} 
%  = \left[ {a}_i, {a}_j^\dagger \right]_{q_{i,j}} 
\end{align}
that the particles acquire an anyonic phase factor $e^{\mathrm{i} \theta}$ under exchange analogous to abelian 2D anyons. 
Here, $\mathrm{sgn}(j)=j/\vert j\vert$ and $\mathrm{sgn}(0)=0$.
These lattice anyons of the so-called anyon Hubbard model
attracted significant attention due to several proposals to realize it in optical lattices filled with ultracold bosons or fermions \cite{KeilmannLanzmichMcCullockRoncaglia2011StatisticallyInducedPhaseTransitionsAndAnyonsIn1DOpticalLattices,GreschnerSantos2015AnyonHubbardModelIn1D,Straeter2016,goerg2019,Clark2018,Cardarelli2016,Dai2017,Lange2017,1DAnyons}
and promising developments of artificial gauge fields and induced phase transitions 
by time-periodic forcing \cite{Bermudez_2015,Greschner2014,Lienhard2020,Lin2009,Lin2011,Rapp2012,Wang2014,Liberto14,Wang2020,Struck2013,Eckardt2017,Arimondo2012,Schweizer2019}. 
In particular, the anyonic exchange phase can be realized 
by an occupation-dependent Peierls-like phase using assisted Raman tunneling \cite{KeilmannLanzmichMcCullockRoncaglia2011StatisticallyInducedPhaseTransitionsAndAnyonsIn1DOpticalLattices,GreschnerSantos2015AnyonHubbardModelIn1D} or periodically modulated lattices \cite{Straeter2016,Clark2018,goerg2019}.
While these lattice anyons faithfully recover the exchange phase from their two-dimensional continuous counterparts, this discrete construction 
side-steps the most interesting aspects of anyon physics:  The continuous 
exchange of two particles.  To our knowledge, the relation between 
1D Leinaas-Myrheim particles, the Kundu model, and the 
experimentally accessible lattice anyons has up to now been an open problem. 
In particular, no corresponding {bosonic} continuum Hamiltonian exists that is $2\pi$-periodic in the anyonic exchange angle.

In this paper, we provide the continuum theory 
of the anyon Hubbard model.  
We thereby obtain an explicit mapping between experiments on one-dimensional bosonic lattices, lattice
theories of anyons, and general theories in the continuum, including Kundu anyons as a special case.
Deriving a naive long wavelength limit is insufficient.
Instead, it is necessary to use the bosonic form of the Hamiltonian and to 
consider all orders of the phase angle. This leads to a {statistically} induced current-density as well as {two-} and three-particle interactions, but as a reward results in the $2\pi$-periodicity in the 
anyonic phase angle even in 1D. 
We emphasize that the bosonic form of the Hamiltonian is experimentally directly accessible, in stark contrast to the purely anyonic description, which encodes the topological character of the exchange
implicitly in the creation algebra or the boundary conditions of the wave functions.  
Additionally, the bosonic form facilitates theoretical calculations, since the anyonic exchange algebra is not preserved under unitary transformations.
We furthermore provide numerical results using the density matrix renormalization group 
(DMRG) algorithm to illustrate how the continuum limit is approached and discuss the implications for the effective low-energy Tomonaga-Luttinger liquid theory.
Nonabelian extensions or unitary braided fusion categories and corresponding diagrammatic equations of anyon theories \cite{Kitaev2006AnyonsInAnExactlySolvableModelAndBeyond,wang2010topological} are interesting prospects for future research and not discussed here.

%%%%%%%%%%%%%%%%%%%%%%%%%%%%%%%%%%%%%%%%%%%%%%%%%%%%%%%%%%%%%%%%
% main text start
%%%%%%%%%%%%%%%%%%%%%%%%%%%%%%%%%%%%%%%%%%%%%%%%%%%%%%%%%%%%%%%%

Our starting point is the anyon Hubbard model on a 1D lattice with $L$ sites \cite{KeilmannLanzmichMcCullockRoncaglia2011StatisticallyInducedPhaseTransitionsAndAnyonsIn1DOpticalLattices}
\begin{align}
\label{eqnAnyonHubbardModel_anyonic}
{H} = -J \sum_{j=1}^{L-1} \left({a}_j^\dagger {a}_{j+1} + \mathrm{h.c.} \right) + \frac{U}{2}\sum_{j=1}^{L} {n}_j({n}_j-1),
\end{align}
where the anyonic operators obey the algebra in \refEq{eqnQuantumbrackets} and  ${n}_j={a}^\dagger_j {a}_j$ is the particle number operator.
Bosons are described by this model at $\theta=0$. 
For $\theta=\pi$ the on-site quantum brackets \refEq{eqnQuantumbrackets} remain bosonic, so that
so-called pseudo-fermions are originally included. Yet, as we show below, the low-density continuum limit captures the behavior of ordinary fermions as well.
The anyon Hubbard model breaks spatial inversion symmetry and time reversal symmetry but obeys a generalized inversion symmetry \cite{Lange2017-1}, that is reduced to the combined action of time reversal and spatial inversion in the below continuum theories, for details, see the Appendix.
By the Jordan-Wigner transformation \cite{KeilmannLanzmichMcCullockRoncaglia2011StatisticallyInducedPhaseTransitionsAndAnyonsIn1DOpticalLattices}
\begin{align}
 {a}_j=e^{i\theta\sum_{k<j}{{n}_k}}{b}_j,
\label{anyon-boson}
\end{align}
the relations in Eq.~(\ref{eqnQuantumbrackets}) can be 
exactly represented by bosonic operators ${b}_j$  using the Hamiltonian
\begin{align}
\label{eqnAnyonHubbardModel_bosonic}
{H} = -J\sum_{j=1}^{L-1} \left( {b}_j^\dagger e^{\I \theta {n}_j}{b}_{j+1}+\mathrm{h.c.}\right) + \frac{U}{2}\sum_{j=1}^{L} {n}_j({n}_j-1).
\end{align}
Here, the hopping depends on the occupation number in form of a Peierls-like factor 
$e^{i\theta {n}_j}$, which plays 
a central role for experimental realizations \cite{KeilmannLanzmichMcCullockRoncaglia2011StatisticallyInducedPhaseTransitionsAndAnyonsIn1DOpticalLattices,GreschnerSantos2015AnyonHubbardModelIn1D, Straeter2016,Clark2018,goerg2019,Schweizer2019}.
Interestingly, the Hubbard interaction ${n}_j({n}_j-1)$ is independent of the anyonic phase {because ${n}_j={a}^\dagger_j{a}_j=b^\dagger_j b_j$}. In the 
continuum limit, it leads to a simple two-body interaction term \cite{Essler2010TheOneDimensionalHubbardModel,Muth2010}, which can be added to the effective interaction terms arising from 
the anyonic exchange derived in the following.

The continuum limit is defined as a process of a systematic renormalization of all terms
when reducing the lattice spacing $d$ with increasing number of sites $L$ 
such that the physical length $l=Ld$ remains finite. 
% Formally, the 
% continuum Hamiltonian becomes
% \begin{align}
% \label{eqnHContinuumDefinition}
% \mathcal{H}_{\text{cont.}} \equiv  
%     \lim_{
%         \parbox{12ex}
%             {\scriptsize \centering $L\rightarrow\infty,$ 
%                 \\ \centering $d\rightarrow 0,Ld=l$
%             }
%         }
%     H \right).
% \end{align}  
Following the procedure of \citeRef{Essler2010TheOneDimensionalHubbardModel}, the bosonic field operator in the continuum is 
\begin{align}
{\Psi}_{B}^{\dagger}(x)=\lim\limits_{d\rightarrow 0}{b}^{\dagger}_{j}/\sqrt{d},
\label{eqnContinuizationTrick}
\end{align} 
which results in the bosonic commutator
\begin{align}
\left[ {\Psi}_{B}(\tilde{x}), {\Psi}_{B}^{\dagger}(x) \right]
% =
% \lim\limits_{a\rightarrow 0}\left[{b}_{i}/\sqrt{a},{b}^{\dagger}_{j}/\sqrt{a}\right]
=\lim\limits_{d\rightarrow 0}\frac{\delta_{i,j}}{d} 
\equiv \delta(x-\tilde{x}).
\end{align} 
Because of the delta-function, it is important that all expressions are normal ordered
before taking the continuum limit in order to avoid divergences.
Furthermore, we expand the bosonic operator as 
\begin{align}
\label{eqnGradientExpansion}
{\Psi}_{B}(x+d)\approx{\Psi}_{B}(x)+d\partial_{x}{\Psi}_{B}(x)+\frac{d^2}{2}\partial_x^2{\Psi}_{B}(x),
\end{align}
since higher orders renormalize to zero in the continuum limit.
In order to keep the full dependence on the anyonic phase angle $\theta$ it is crucial
to express the Peierls-like factor in normal ordered form, which is possible
to all orders according to \cite{Blasiak2007CombinatoricsandBosonnormalorderingAgentleintroduction}
\begin{align}
e^{i\theta{n}_{j}}=\sum_{q=0}^{\infty}\frac{(i\theta)^q}{q!}{n}_{j}^{q}=\sum_{q=0}^{\infty}\frac{(i\theta)^q}{q!}\sum_{m=0}^{q}S(q,m)({b}^{\dagger}_{j})^{m}({b}_{j})^{m},
\label{peierls}
\end{align}
where $S(q,m)$ are the Stirling numbers of second kind \cite{Sexton2012AbramowitzandStegunAResourceforMathematicalDocumentAnalysis}.
Taking the continuum limit according to \refEq{eqnContinuizationTrick} and \refEq{eqnGradientExpansion},  we observe that each 
operator ${b}_{i}$ carries a factor of $\sqrt{d}$ and each derivative a factor of $d$, which yields an overall scale $Jd^2 =\hbar^2/2 m_{\rm eff}$ that is
set to unity in the following.  By neglecting higher powers in $d$ and resumming the 
Peierls-like factor in \refEq{peierls}, we finally obtain  
the Hamiltonian in the continuum limit
\begin{align}\label{fullH}
\mathcal{H}_{\mathrm{cont.}} = \int dx\phantom{a}\Bigl[ \mathcal{H}_{\mathrm{kin}}(x)+\mathcal{H}_{\mathrm{int}}(x)+\mathcal{H}_{\mathrm{J}}(x)\Bigr],
\end{align}
with 
\begin{align}
\label{eqnbosonicfieldtheory}
\mathcal{H}_{\mathrm{kin}}(x)=&\partial_{x}{\Psi}_B^{\dagger}(x)\partial_{x}{\Psi}_B(x)\nonumber,\\
\mathcal{H}_{\mathrm{int}}(x)=& \left( V_{2}(\theta)+c \right):\rho_B^2(x):+V_{3}(\theta):\rho_B^3(x):,\\
\mathcal{H}_{\mathrm{J}}(x)=& V_J(\theta):\rho_B(x){J}_B(x): \nonumber,
\end{align}
where\phantom{a}\mbox{${J}_B(x)=-i\left({\Psi}_B^{\dagger}(x)\partial_{x}{\Psi}_B(x)-\mathrm{h.c.}\right)$} is the current operator, \mbox{$\rho_B(x)={\Psi}_B^{\dagger}(x){\Psi}_B(x)$} the density operator and $:\bullet:$ denotes normal order. The parameter $c$ is related to the Hubbard interaction $U = 2 J d c$, and we have furthermore dropped the energy dependence on the total density $\rho_0=N/Ld$, since
the particle number $N$ is a conserved quantity.
%adjusted the chemical potential to vanish.
The coupling constants of the theory are
\begin{align}
V_J(\theta)=&\ \mathrm{Im}\left[\tilde{V}_{2}(\theta)\right]=-\sin(\theta),
\\
V_{2}(\theta)=& -2d^{-1}\mathrm{Re}\left[\tilde{V}_{2}(\theta)\right]={2d^{-1}\left[1-\cos(\theta)\right]}, \label{divergence}
\\
V_{3}(\theta)=&-\mathrm{Re}\left[\tilde{V}_{2}(\theta)^2\right]=1+2\cos(\theta)-\cos(2\theta),
\end{align}
with $\tilde{V}_{2}(\theta)=1-e^{i\theta}$.
The two-body interaction $V_{2}(\theta)$ and the rescaled Hubbard interaction $c=U/(2Jd)$ strongly renormalize
when $d\rightarrow0$ if $\theta\neq0$, which is known to occur also for the 
ordinary Fermi-Hubbard and Bose-Hubbard interaction $U$ in the continuum limit
\cite{Cazalilla2003,Essler2010TheOneDimensionalHubbardModel}.
In the Hubbard model this 
can be resolved by rescaling 
$U$ with the overall density, leading to a renormalized theory 
which recovers the so-called Tonks-Girardeau limit with infinitely strong interactions for low densities
\cite{Tonks1936TheCompleteEquationofStateofOneTwoandThreeDimensionalGasesofHardElasticSpheres,Girardeau1960}.  However, in the anyonic theory, the angle $\theta$ cannot simply be rescaled or renormalized without changing the anyonic phase angle, which would violate the topological character. 
As we see below, and justified by our numerical results, the dependence on the lattice spacing $d$ allows us to define regularized coupling constants, which are fixed by experimental parameters.
Moreover, we recognize a current-density interaction potential $V_J$ and a
three-body interaction $V_3$, so that the full theory cannot be derived 
by only considering two-particle scattering processes.

We furthermore observe that 
the resulting bosonic model in Eq.~(\ref{fullH}) 
has the same structure as the integrable Kundu model \cite{Kundu1999ExactSolutionOfDoubleDeltaFunctionBoseGasThroughAnInteractingAnyonGas} albeit with 
different coupling constants.  %In particular, 
The 
Kundu model is described by 
\begin{align}
V_J^{\rm Kundu}(\theta)=& -\theta, \label{kundu1} \\
V_{2}^{\rm Kundu}(\theta)=&\ \theta^2 \int dx\hspace{0.1cm} \delta^2(x) + c,
\\
V_{3}^{\rm Kundu}(\theta)=&\ \theta^2. \label{kundu3}
\end{align}
Up to second order in $\theta$, the parameters in Eqs.~(\ref{kundu1})--(\ref{kundu3}) recover the 
% which describes a possible additional interaction of the equivalent 
anyonic Kundu model  \cite{Kundu1999ExactSolutionOfDoubleDeltaFunctionBoseGasThroughAnInteractingAnyonGas}
\begin{align}
{\mathcal{H}}_{\mathrm{K}}=&\displaystyle{\int_{-\infty}^{\infty}dx\phantom{a}} \partial_{x}{\Psi}_A^{\dagger}(x)\partial_{x}{\Psi}_A(x)\nonumber
\\
+&\displaystyle{\int_{-\infty}^{\infty}dx\phantom{a}}{c}\hspace{0.1cm} {\Psi}_A^{\dagger}(x){\Psi}_A^{\dagger}(x){\Psi}_A(x){\Psi}_A(x).
\end{align}
Here, the anyonic operators ${\Psi}_A$ are related to the bosonic ones ${\Psi}_B$ by the continuum version of the Jordan-Wigner transformation \cite{Kundu1999ExactSolutionOfDoubleDeltaFunctionBoseGasThroughAnInteractingAnyonGas}
\begin{align}
{\Psi}_A(x)=e^{i\theta\int_{-\infty}^{x}{n}(y)dy}{\Psi}_B(x),
\label{kundulast}
\end{align}
where ${n} = {\Psi}_B^{\dagger}{\Psi}_B^{\phantom{\dagger}}$.
This relation naively looks like a straight-forward 
continuum limit of Eq.~(\ref{anyon-boson}), but such an 
approximation does not capture the full topological 
character since the crucial symmetry $\theta\to \theta + 2\pi$ is lost.

By comparing the coupling constants, we see that the Kundu model ${\mathcal{H}}_{\mathrm{K}}$ 
corresponds to the special case of small anyon angles $\theta$ 
in the %full
continuum model $\mathcal{{H}}_{\text{cont.}}$ in 
Eq.~(\ref{fullH}).  
Moreover, 
the singular behavior of a double delta function is replaced by
the $1/d$ divergence of $V_2$ in Eq.~(\ref{divergence}).
Therefore, our derived continuum model is more general and introduces a well-defined limiting procedure how the controversially discussed  \cite{Girardeau2006AnyonFermionMappingandApplicationstoUltracoldGasesinTightWaveguides,PatuKorepinAverin2008} double delta function in the original 
Kundu model must be interpreted. Namely, we find that one has to extrapolate the experimental results when lowering the lattice spacing $d$ towards zero while keeping $L d$ finite.
Obviously, changing $d$ directly is difficult in an optical lattice, but 
it is possible to reach this limit by noting that the density $\rho_0=N/Ld$ remains finite in the continuum limit. For a given density $\rho_0$,
the continuum limit can then be effectively achieved by extrapolating $d = N/L\rho_0 \to 0$ by 
systematically lowering the number of particles per site $N/L$.

We illustrate this procedure by means of a numerical experiment using the 
%density matrix renormalization group 
DMRG algorithm \cite{White1}, which is a powerful tool to analyze the properties of the
proposed continuum theory in Eq.~(\ref{fullH}).  
We simulate non-interacting
anyons by bosons with an occupation-dependent hopping in the continuum limit of \refEq{eqnAnyonHubbardModel_bosonic} with $U=0$
using up to $500$ DMRG states in finite systems with 
fixed boundary conditions at the edges.
For the values of $\theta$ and the
relatively low densities used for the simulations below, we find that the numerical restriction to at most two bosons per site gives good results.
It is well-established that the local density can be considered as a
convenient observable to analyze the interaction strength in 1D \cite{Straeter2016,friedel1,friedel2,Rommer2000,LLBox,soeffing2009}, since
characteristic density oscillations develop near the edges due to collective 
modes, which ultimately are related to Friedel oscillations in the fermion limit \cite{friedel}.  An interacting  bosonic gas gradually develops density oscillations near edges \cite{Muth2010} that grow with increasing interactions and with decreasing densities analogous to the Fermi-Hubbard model \cite{soeffing2009}.

\begin{figure}
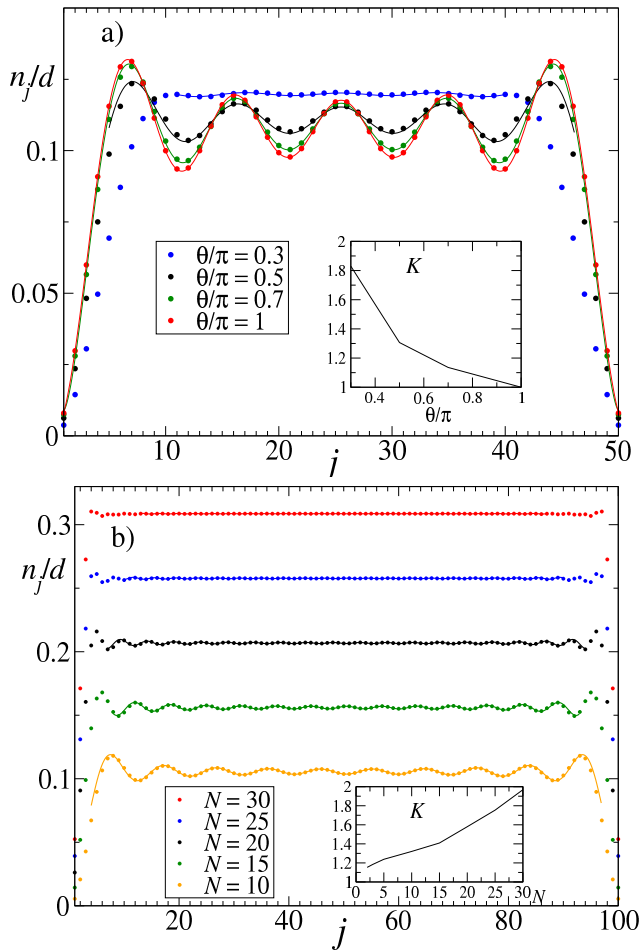

%\begin{center}
%\subfloat[\label{figDMRGVariedTheta}]
%{\mbox{
\hspace*{-2.5ex}
\includegraphics[width=.96\columnwidth]{fixed-n_2_.eps}
% \\~\\
%\\
%\subfloat[\label{figDMRVariedDensity}]
{\includegraphics[width= .95\columnwidth]{fixed-theta_2_.eps}}
%\end{center}
\caption{\label{figDMRG} DMRG results for the local particle density $n_j/d$.
{(a)} For a chain of $L=50$ sites with $N=5$ anyons with different $\theta$.
{(b)} For $\theta=0.5\pi$ and a chain length $L=100$ the Friedel oscillations increase with decreasing particle number $N$. Solid lines are fits to powerlaw-decaying oscillations in Eq.~(\ref{fit}), from which the Luttinger parameters $K$ are extracted (insets).}
\end{figure}

In \refFig{figDMRG}a, we see that the corresponding density oscillations in a non-interacting
anyon gas grow with $\theta$, which plays the role of an effective interaction as expected.
Using $N=5$ particles and $L=50$ sites, 
the densities gradually build up with increasing values of $\theta$.
Keeping $\theta=0.5\pi$ fixed in Fig.~\ref{figDMRG}b, we observe how the characteristic oscillations 
steadily increase with lower densities.
 We see that choosing a small  
but finite density $N/L$ creates a natural cutoff similar to choosing a finite $d$ in the bosonization procedure to renormalize diverging terms in impurity problems \cite{VonDelftSchoellerBosonizationForBeginnersRefermionizationForExperts}.

In order to understand the oscillations on a more quantitative level, 
we derive the corresponding Tomonaga-Luttinger liquid (TLL) theory \cite{Giamarchi2004}
describing the low-energy excitation of interacting bosons.  Using 
the phase-density
representation of the bosonic fields from the harmonic fluid approach
\cite{Cazalilla2004Bosonizingonedimensionalcoldatomicgases,Haldane1981EffectiveHarmonicFluidApproachtoLowEnergyPropertiesofOneDimensionalQuantumFluids} for 
the interactions of Eq.~(\ref{eqnbosonicfieldtheory}),
we immediately arrive at a TLL Hamiltonian together with a special 
parity-breaking interaction $\Delta$ (see Appendix)
\begin{equation}
\mathcal{H}_\mathrm{cont.}\approx\frac{u}{2\pi}{\displaystyle \int dx} \left[\frac{1}{K}(\partial_{x}\phi)^2+K(\partial_{x}\Theta)^2+\Delta\partial_{x}\phi\partial_{x}\Theta\right],
\end{equation}
where energy shifts due to conserved quantities 
(total current and density) have again been omitted (see Appendix).
The dependence of the Luttinger parameter $K$, the Fermi-velocity $u$ and the 
interaction $\Delta$ 
on the parameters $c$, $V_2$, $V_3$, and $V_J$ can be read off to lowest order
$\pi^2/K^2 = \left(V_2 +c\right)/\rho_0 + 3 V_3$, $u=2\pi \rho_0/K$, and 
$\Delta  = 8 \rho_0 V_J/u$. But note that the
corresponding expressions from the harmonic fluid approach
are known to be only reliable for very small coupling constants \cite{Cazalilla2004Bosonizingonedimensionalcoldatomicgases,haueslertomonagaluttingerparametersforquantumwires2002}, i.e., for small $\theta$ and $c$.
Therefore, other methods are required to obtain a 
quantitative estimate of $K$ for all $\theta$ as discussed below.
The non-trivial 
interaction $\Delta$ has interesting physical consequences. 
Under spatial parity inversion, the fields transform $\partial_x\phi\to \partial_x\phi$ and $\partial_x\Theta\to -\partial_x\Theta$ \cite{eggert1992}, 
so this interaction can be traced back to the parity breaking nature of
the anyonic description.  
However, the $\Delta$-term is
not affected by the Luttinger rescaling of the fields $\phi\to \sqrt{K} \phi$ and
$\Theta\to \Theta/\sqrt{K}$ and 
becomes diagonal using the known mode expansions in terms of collective density waves
\cite{Cazalilla2004Bosonizingonedimensionalcoldatomicgases} as shown in 
the Appendix, 
so no additional transformation is required.  
Therefore, the vaccum and the excitations of a TLL
are preserved, but the interaction $\Delta$
leads to different velocities and energies of left- and right-moving density 
excitations $\epsilon_q=\left( u + \Delta \sgn(q)/2\right)\vert q\vert$, which becomes relevant
for time-dependent correlations (see Appendix).

In order to provide a quantitative estimate of the Luttinger parameter $K$
for all $\theta$, the numerical data in Fig.~\ref{figDMRG} prove useful.
It is known that the long-range decay of the
oscillations away from the edges 
is governed by a powerlaw \cite{friedel1,friedel2}, where the exponent, for spinless models, 
is the Luttinger parameter $K$  \cite{Rommer2000}.
In particular, the oscillations in the local density of a 
finite-size TLL follow the analytic 
expression \cite{Cazalilla2004Bosonizingonedimensionalcoldatomicgases,soeffing2009,calabrese}
\begin{equation}
\label{fit}
n_j/d \approx \rho + A \cos{  \left[2  \rho\pi (j d- \ell/2)\right]}\left(\ell \sin \frac{\pi j d}{\ell}\right)^{\!\!-K}\!\!\!,
\end{equation}
where $\rho$ is the average density near the middle of the chain $\ell/2 \equiv (L+1)d/2$. 
For the ground state expectation value of $n_j$ the parity-breaking interaction $\Delta$ 
does not contribute (see Appendix).
As shown in Fig.~\ref{figDMRG}, Eq.~(\ref{fit}) describes the local density very well 
for all $\theta > 0$ and results in
a non-trivial exponent $K$ which approaches unity 
for small densities and $\theta \to \pi$. The fits are spatially limited by a 
cut-off distance from the edges, which indicates the range of validity of
the TLL theory.    
Notably, the cut-off distance increases with the Luttinger parameter $K$. It hence becomes
increasingly difficult to describe weaker oscillations, but the data
is still consistent with the expected behavior of free bosons $K\to \infty$ for 
$\theta \to 0$, see inset of Fig.~\ref{figDMRG}a. In the opposite limit 
of $\theta =\pi$, the solid line in Fig.~\ref{figDMRG}a 
is given by the analytic expression for 
free fermions \cite{soeffing2009} over the entire system,
which means that the pseudo-fermions are
well described by ordinary fermions with $K=1$ in this case \cite{GreschnerSantos2015AnyonHubbardModelIn1D,Straeter2016}. 
The TLL parameter in the insets of  Fig.~\ref{figDMRG} determines relevant long-range 
correlations \cite{Giamarchi2004} including energy-dependent quantities like
the local density of states \cite{STM}.  The numerical data therefore not only confirms
the stability of a TLL ground state but also predicts the dominant correlations of the continuum anyon model quantitatively as a function of $\theta$ and $\rho_0$.
In experimental setups, the confinement is commonly given by a harmonic trap, which also 
gives rise to density oscillations \cite{Muth2010} described by  
corresponding fit functions \cite{harmonic}. Thus, experimental measurements of
the local densities \cite{Vogler2014} in optical traps can be used to extract these characteristic exponents.

%%%%%%%%%%%%%%%%%%%%%%%%%%%%%%%%%%%%%%%%%%%%%%%%%%%%%%%%%%%%%%%%
% end text start
%%%%%%%%%%%%%%%%%%%%%%%%%%%%%%%%%%%%%%%%%%%%%%%%%%%%%%%%%%%%%%%%
In conclusion, capturing the essential feature of a 2D anyonic exchange phase by 1D anyons on a lattice inspired our development 
of a continuum theory of 1D anyons in terms of interacting bosons, which
keeps the topological character of 
abelian anyonic exchange in contrast to previously discussed Hamiltonians in 1D. 
The representation of anyons as normal ordered bosons with modified hopping is crucial
when taking the continuum limit. We take all orders of the anyonic phase into account, resulting in a Hamiltonian that includes current-density as well as two- and three-particle bosonic interactions with $2\pi$-periodic coefficients in the anyonic phase angle.
The description of 1D continuum anyons, 
%in one dimension 
that captures this topological hallmark,
solves an open problem and the theory extends the two-dimensional 
concept of a $2\pi$-periodic anyonic exchange phase to 1D.
The known Kundu model is contained in the limit of a vanishing statistical angle.
Our work therefore uncovers a unifying, physically motivated continuum theory of one-dimensional abelian anyons that derives from the original idea of an anyonic exchange phase in 2D.
We thereby connect recent experiments on ultracold atomic gases to the seminal considerations of 
Leinaas and Myrheim  as well as Wilczek,
% \cite{Wilczek1982QuantumMechanicsOfFractionalSpinParticles,LeinaasMyrheim1977TheoryOfIdenticalParticles}
which inspired the name `anyon' \cite{Wilczek1982QuantumMechanicsOfFractionalSpinParticles}.
In physical systems, we quantitatively predict the appearance of characteristic 
density oscillations, which reflect the TLL correlations. 
Our studies also show that a hallmark of anyonic physics are 
different velocities of left- and right-moving density excitations,
which could be observable in dynamic experiments on the structure factor or time-dependent correlators (see Appendix).

\begin{acknowledgments} 
T.P.\ and M.T.\  acknowledge support {
by the Cluster of Excellence ``CUI: Advanced Imaging of Matter'' of the Deutsche Forschungsgemeinschaft (DFG) -- EXC 2056 -- project ID 390715994. M.B., K.J., A.P., and S.E. were funded by the DFG
% the Deutsche Forschungsgemeinschaft (DFG, German Research
% Foundation)
-- Project-ID 277625399 -- TRR 185 via the SFB/Transregio 185: ``OSCAR -- Open System Control of Atomic and Photonic Matter''.} Furthermore, A.P. thanks for funding from Volkswagen Foundation  under project number 86896.
\end{acknowledgments}

\section{Appendix}
In this Appendix, we summarize the harmonic fluid approach and 
and mode expansions.
We use it to
derive the Tomonaga-Luttinger liquid (TLL) theory of the anyon model and the time-dependent density-density correlator is examplarily
 calculated to show the effect of $\Delta$..
%In order to be self-contained, we repeat the necessary formulas from the main text.

\subsection{Harmonic Fluid Bosonization}

\label{sctnHarmonicFluidBosonization}

{For obtaining} the low-energy properties of the continuum model, we follow the harmonic fluid approach \cite{Cazalilla2004Bosonizingonedimensionalcoldatomicgases,Haldane1981EffectiveHarmonicFluidApproachtoLowEnergyPropertiesofOneDimensionalQuantumFluids},
 where the
density operator and the creation operator are approximated in terms of fluctuations of a bosonic field $\phi(x)$ and the dual field $\Theta(x)$
using a number-phase representation
\begin{align}
\rho_B(x)&\approx\left[\rho_0-\frac{1}{\pi}\partial_{x}\phi(x)\right]\sum_{m=-\infty}^{\infty} e^{2im[\pi\rho_0x-\phi(x)]}
\\
\label{harmonicfluidcreationoperator}
	\Psi_B^{\dagger}(x)&\approx\sqrt{\rho_0-\frac{1}{\pi}\partial_{x}\phi(x)}\sum_{m=-\infty}^{\infty} e^{2im[\pi\rho_0x-\phi(x)]} e^{-i\Theta(x)},
\end{align}
with the commutation relation
%and $\Theta(x)$ is the dual field,
%The method is based on a number-phase representation of the field operators in terms of the dual fields $\phi,\Theta$
\begin{align}
	\left[\phi(x),\partial_{y}\Theta(y)\right]=i\delta(x-y).
\end{align}
%describing number and phase fluctuations of the fluid, respectively.
We neglect higher orders of the density fluctuations in the square root expansion as well as contributions from higher harmonics, which amounts to only considering small density fluctuations and neglecting lattice effects, respectively  \cite{Cazalilla2004Bosonizingonedimensionalcoldatomicgases}.
Employing subsequently the bosonization dictionary from Ref.~\citeRef{Cazalilla2004Bosonizingonedimensionalcoldatomicgases},
we obtain for the kinetic part
\begin{align}
	\partial_{x}\Psi_B^{\dagger}(x)\partial_{x}\Psi_B(x)&\approx \rho_0(\partial_{x}\Theta)^2,
\end{align}
while the interaction terms in Eq.~(11) of the main text can be expressed as
\begin{align}
	:\rho_B(x)J_B(x):\phantom{a}\approx &-2\rho^2_0\ \partial_{x}\Theta +\frac{4\rho_0}{\pi}\partial_{x}\Theta\partial_{x}\phi,
\end{align}
\begin{align}
	% 	&V_{3}(\theta)
	: \rho^{3}_B(x):\phantom{a}\approx
% V_{3}(\theta)
	&-\rho^{2}_{0}\frac{3}{\pi}\partial_{x}\phi+\rho_{0}\frac{3}{\pi^2}(\partial_{x}\phi)^2,
\\
% &\left[V_{2}(\theta)-3V_{3}(\theta)\right]
:\rho^{2}_B(x):\phantom{a}
% \\
%
\approx
% \left[V_{2}(\theta)-3V_{3}(\theta)\right]
&-\frac{2}{\pi}\rho_{0}\partial_{x}\phi+\frac{1}{\pi^2}(\partial_{x}\phi)^2.
\end{align}
After integration the terms linear in the fields $\partial_x \phi$ and $\partial_x \Theta$ give
contributions to the energy due to the total particle number $N$ and the total current $J$ in
the system \cite{Haldane1981EffectiveHarmonicFluidApproachtoLowEnergyPropertiesofOneDimensionalQuantumFluids, Cazalilla2004Bosonizingonedimensionalcoldatomicgases}.  Since these are conserved quantities, we will not consider
those energy shifts in the following.
Collecting all bilinear terms, the low-energy field theory can be expressed in terms of a TLL
Hamiltonian
\begin{equation} \label{ll}
	\mathcal{H}_{\mathrm{cont.}}\approx\frac{u}{2\pi}{\displaystyle \int dx} \left[\frac{1}{K}(\partial_{x}\phi)^2+K(\partial_{x}\Theta)^2+\Delta\partial_{x}\phi\partial_{x}\Theta\right],
%\nonumber \\
%	+&{\displaystyle \int dx}\left[\epsilon_N\partial_{x}\phi+\epsilon_J\partial_{x}\Theta,\right],
\end{equation}
with the Luttinger parameter $K$
\begin{align}
	K=\sqrt{\frac{\pi^2\rho_0}{V_{2}(\theta)+c+3\rho_{0}V_{3}(\theta)}},
	\end{align}
\\
the Fermi velocity $u$
\begin{align}
u=&2 \pi \rho_0/K,
\end{align}
and the parity-breaking interaction $\Delta$
\begin{align}
	\Delta=&8 \rho_0 V_J(\theta)/u.
\end{align}
	%u=&2\sqrt{\rho_0\left[V_{2}(\theta)+3\rho_0V_{3}(\theta)\right]},

	%\epsilon_J=&-2 \rho^2_0V_J(\theta), \\
% \epsilon_N=&-\frac{2\rho_0}{\pi}V_2(\theta)+\frac{3\rho_{0}^2}{\pi}V_3(\theta), \\
	%\gamma=&\frac{\pi}{u}\left[4V_J(\theta)\frac{\rho_0}{\pi}\right].

\subsection{Mode expansion}

In this section, we show how the mode expansion diagonalizes the TLL Hamiltonian, derive the different velocities and energies of left- and right-moving excitations, and give a time-dependent bulk-density correlation function.

For a lattice on a ring we need to consider periodic or twisted boundary conditions.
Due to the Jordan-Wigner string, it is known that periodic boundary conditions
for the anyonic operators lead to twisted boundary conditions of the bosons
$\Psi_B(L)=\Psi_B(0)e^{i\theta {N}}$ \cite{Cazalilla2004Bosonizingonedimensionalcoldatomicgases}, %formerly [33], Kundu
and vice versa.
%If the anyonic lattice Hamiltonian obeys periodic boundary conditions its bosonic pendant obeys twisted ones, where the twist angle depends on the total number of particles due to counting ambiguities in the Jordan-Wigner string at the boundary, i.e., $\hat{\Psi}_B(L)=\hat{\Psi}_B(0)e^{i\theta\hat{N}}$ [33]. %\cite{Kundu1999ExactSolutionOfDoubleDeltaFunctionBoseGasThroughAnInteractingAnyonGas}.
By shifting the total current $J$ accordingly  $\delta J+\frac{\theta N}{\pi}=J$,
such boundary twists can be incorporated into
the usual bosonic mode decomposition, according to \cite{Cazalilla2004Bosonizingonedimensionalcoldatomicgases}
\begin{align}
&\phi(x)=\phi_{0}-\frac{\pi Nx}{L}-\frac{i}{2}\sum\limits_{q\neq0}\sqrt{\frac{2\pi K}{L\vert q\vert}}\frac{q}{|q|}
%\mathrm{sgn}(q)
e^{-iqx}\left[\beta^{\dagger}_{q}+\beta_{-q}\right]
\\
&\Theta(x)=\Theta_{0}+\frac{\pi\delta Jx}{L}+\frac{i}{2}\sum\limits_{q\neq 0}\sqrt{\frac{2\pi}{ KL\vert q\vert}}e^{-iqx}\left[\beta^{\dagger}_{q}-\beta_{-q}\right],
\end{align}
%by shifting the total current $J$ accordingly, i.e., $\delta\hat{J}+\frac{\theta N}{\pi}=\hat{J}$.
where $\beta^{\dagger}_q$ creates a collective wave with wave vector $q=\frac{2\pi m}{L}$.
The conserved current $ J$ and particle number $ N$
%Here $\delta\hat{J},\delta\hat{N}$
can be interpreted as action-angle variables dual to $\phi_{0},\Theta_{0}$, which  are used to label the topological excitations of the system.  As before we omit their dependence on the
total energy of the system.
Insertion of the mode decomposition into the TLL in \refEq{ll}
 leads then to the modified diagonal form
\begin{equation}
\label{eqnmodehamiltonianperiodic}
\mathcal{H}_{\mathrm{cont.}}\approx \sum_{q\neq 0}
\left[u + \Delta \sgn(q)/2\right]
\vert q\vert
\beta^{\dagger}_q\beta_q  .
\end{equation}
The impact of the interaction $\Delta$ is clearly visible. It changes
the velocity by $\Delta/2$ depending on the sign of $q$,
%In Eq.~(\ref{eqnmodehamiltonianperiodic}), the velocities of the finite momentum components are asymmetrically shifted by $\gamma$,
implying the breaking of parity symmetry $q\leftrightarrow -q$. 
% inherent to such anyonic models [34]. %\cite{KeilmannLanzmichMcCullockRoncaglia2011StatisticallyInducedPhaseTransitionsAndAnyonsIn1DOpticalLattices}.
%Furthermore the zero modes of the density and current degrees of freedom are coupled by the presence of $\Delta$, interpretable as momentum shift of the Hamiltonians ground state.
We therefore predict different velocities and energies for left- and right-moving excitations.
However, the eigenstates of the Hamiltonian are unchanged by $\Delta$.  In particular,
the ground state
expectation value of the local density $n_j$ discussed in the main text therefore takes the
usual TLL form, which can be used to extract $K$. Time-dependent correlators on the other hand are modified as we examplarily show for the bulk density-density correlation function including all higher harmonics \cite{Cazalilla2004Bosonizingonedimensionalcoldatomicgases}, i.e.,
\begin{align}
\label{densitycorrelator}
&\langle \rho_B(x,t)\rho_B(0,0) \rangle=\rho^2_0
\\
&-\frac{K}{2(\rho_0\pi)^2}\left[\frac{1}{2\bigl\vert x-u\left(1-\Delta/2\right)t\bigr\vert^2}+\frac{1}{2\bigl\vert x+u\left(1+\Delta/2\right)t\bigr\vert^2}\right] \nonumber
\\
&+\rho^2_0\sum_{m>0}\alpha_m \frac{\cos(2\pi m\rho_0x+\delta_m)}{\Bigl(\rho_0\bigl\vert x-u[1-\Delta/2]t\bigr\vert\bigl\lvert x+u[1+\Delta/2]t\bigr\rvert\Bigr)^{m^2 K}}.\nonumber
\end{align}
Here, $\alpha_m, \delta_m$ denote non-universal numerical constants related to the short wave-length structure of the Hamiltonian.

\subsection{Symmetries}
In this section, we explain the symmetry properties of the lattice Hamiltonian, the low-density theory, as well as the corresponding Tomonaga-Luttinger liquid theory of the main text. Here, the focus resides on the spatial parity and time-reversal symmetry, whereas particle number conservation and boundary-dependent symmetries are not discussed.
\\
The lattice Hamiltonian $H$ in its bosonic form, Eq.~(4) of the main text, includes occupation number-dependent hopping processes that breaks time-reversal symmetry $\mathcal{T}$ and spatial parity $\mathcal{P}$ for all $\theta\neq0$ or $\pi$. Here, the antiunitary time-reversal symmetry is represented by complex conjugation $\mathcal{T} = \mathcal{K}$ and the unitary parity reversal by $\mathcal{P}^{\dagger}b_j\mathcal{P}=b_{L-j+1}$, i.e., by mirroring the lattice sites. The absence of the $\mathcal{P}$ and $\mathcal{T}$ symmetry is a major characteristics of these anyons, in contrast to bosonic Hubbard models that obey these symmetries.
% The parity/time transformed Hamiltonians  $K^{\dagger}HK,P^{\dagger}HP,P^{\dagger}K^{\dagger}HPK$ are not invariant under the respective transformation and correspond to different conventions of the anyonic exchange phase $e^{\pm i\theta}$ as well as different embeddings in the underlying two-dimensional space [74].
%
\\
Ref.~\cite{Lange2017-1} finds an antiunitary inversion symmetry $G = U \mathcal{P}\mathcal{T}$, where  $U=e^{-i\frac{\theta}{2}\sum_jn_j(n_j-1)}$ that leaves the bosonic form of the anyonic lattice model invariant, Eq.~(4) of the main text. %$\left[G,H\right]=0$ .
This symmetry acts on the anyonic operators according to $G^{\dagger}a_jG=a_{L-j+1}e^{-i\theta N}$, i.e., as a phase shifted spatial parity inversion. $G$ furthermore is a hidden symmetry of the lattice Hamiltonian in anyonic formulation, Eq.~(2) of the main text.
\\
Performing the continuum limit of the main text in Eq.~(6), the unitary transformation $U$ above reduces to a global phase shift, i.e.,
\begin{align}
\label{specialinversionsymmetryvanishing}
U_{\mathrm{cont.}}=\lim_{d\to 0} e^{-i\frac{\theta}{2}\int dx \left( d \rho_B^2(x)- \rho_B(x) \right)} =  e^{i\frac{\theta N}{2}},
\end{align}
which transforms particle-number conserving Hamiltonian trivially.
Hence, the continuum Hamiltonian $\mathcal{H}_{\mathrm{cont}}$ in Eq.~(10) of the main text, is necessarily $\mathcal{P}\mathcal{T}$ symmetric, manifested in the current-density coupling $\mathcal{H}_{\mathrm{J}}\propto \rho_B(x) J_B(x)$.
The bosonic current $J_B(x)$ changes its direction under time-reversal as well as under spatial inversion, i.e., $x\rightarrow -x$, as both transformations reverse the momenta of the particles, $k\rightarrow -k$. Hence already the combination of both parity and time-reversal becomes a symmetry in the continuum limit.
% Thus we can formally interpret the arbitrary choice of the exchange phase with the preferential direction of the current that flows through the system.
\\
\\
The TLL in \refEq{ll} (Eq.~(20) of the main text) obeys the same symmetries as the low-density theory. Spatial parity inversion transforms the dual fields according to $\partial_x\phi\rightarrow\partial_x\phi$ and $\partial_x\Theta\rightarrow-\partial_x\Theta$, as stated in the main text. Time-reversal results in $\phi\rightarrow\phi$ and $\Theta\rightarrow-\Theta$ , as can be inferred from \refEq{harmonicfluidcreationoperator}. Hence, parts of the Hamiltonian reverse their sign under both respective operators, but the Hamiltonian is invariant under a consecutive application.

\bibliographystyle{apsrev4-1}
%\bibliographystyle{mystyle}
%\bibliography{Bib2020_1DAnyonTheory}

\end{document}